\documentclass[a4paper,superscriptaddress,aps,preprintnumbers,amsmath,amssymb,sort&compress,nofootinbib,apj]{revtex4}

\usepackage{amssymb,amsmath,microtype}
\usepackage{graphicx}
\usepackage[normalem]{ulem}
\usepackage{mathtools,xparse}
\usepackage{latexsym}

\usepackage{color}

\usepackage{amsmath}
\usepackage{gensymb}

\begin{document}
\title{Tibet's Ali: A New Window to Detect the CMB Polarization}

\author{Yong-Ping Li\footnote{E-mail: liyp@ihep.ac.cn}}
\affiliation{Theoretical Physics Division, Institute of High Energy Physics (IHEP), Chinese Academy of Sciences, 19B Yuquan Road, Shijingshan District, Beijing 100049, China}
\affiliation{University of Chinese Academy of Sciences, Beijing, China}

\author{Yang Liu}
\affiliation{Theoretical Physics Division, Institute of High Energy Physics (IHEP), Chinese Academy of Sciences, 19B Yuquan Road, Shijingshan District, Beijing 100049, China}
\affiliation{University of Chinese Academy of Sciences, Beijing, China}

\author{Si-Yu Li}
\affiliation{Key Laboratory of Particle Astrophysics,  Institute of High Energy Physics (IHEP), Chinese Academy of Sciences, 19B Yuquan Road, Shijingshan District, Beijing 100049, China}

\author{Hong Li}
\affiliation{Key Laboratory of Particle Astrophysics,  Institute of High Energy Physics (IHEP), Chinese Academy of Sciences, 19B Yuquan Road, Shijingshan District, Beijing 100049, China}

\author{Xinmin Zhang}
\affiliation{Theoretical Physics Division, Institute of High Energy Physics (IHEP), Chinese Academy of Sciences, 19B Yuquan Road, Shijingshan District, Beijing 100049, China}
\affiliation{University of Chinese Academy of Sciences, Beijing, China}

\begin{abstract}
The Cosmic Microwave Background (CMB) Polarization plays an important role in current cosmological studies.
CMB B-mode polarization is the most effective probe to primordial gravitational waves (PGWs) and a test of the inflation as well as other theories of the early universe such as bouncing and cyclic universe.
So far, major ground-based CMB polarization experiments are located in the southern hemisphere.
Recently, China has launched the Ali CMB Polarization Telescope (AliCPT) in Tibetan Plateau to measure CMB B mode polarization and detect the PGWs in northern hemisphere. AliCPT include two stages, the first one is to build a telescope at the 5250m site (AliCPT-1) and the second one is to have a more sensitive telescope at a higher altitude of about 6000m (AliCPT-2).
In this paper, we report the atmospherical conditions, sky coverage and the current infrastructure associated with AliCPT.
We analyzed the reanalysis data from MERRA-2 together with radiosonde data from the Ali Meteorological Service and found that the amount of water vapor has a heavy seasonal variation and October to March is the suitable observation time.
We also found 95/150 GHz to be feasible for AliCPT-1 and higher frequencies to be possible for AliCPT-2.
Then we analyzed the observable sky and the target fields, and showed that Ali provides us a unique opportunity to observe CMB with less foreground contamination in the northern hemisphere and is complementary to the existed southern CMB experiments.
Together with the developed infrastructure, we point out that Ali opens a new window for CMB observation and will be one of the major sites in the world along with Antarctic and Atacama.

\end{abstract}
\maketitle

\section{Introduction}
Precise measurements of CMB are playing a more and more important role in cosmology nowadays. In the past 20 years, CMB experiments have provided us rich information for understanding the origin and evolution of our universe. Recently, CMB polarization observations have became the study hotspot in the field of particle physics and cosmology. Being created at very early universe, the primordial gravitational waves left unique imprint on the B-mode CMB polarization\cite{Seljak:1996gy}\cite{Kamionkowski:1996zd}. Moreover, CMB polarization can also provide powerful test on the fundamental physics such as CPT symmetry\cite{Li:2008tma}\cite{Zhao:2014yna}\cite{Zhao:2015mqa}\cite{Li:2015vea}. Compared to satellite projects, ground-based experiments offers a more economic and scalable way to detect B-mode signal at high precision.

For ground-based experiments, the atmosphere is a big issue. Transmission in the microwave band depends heavily on atmospheric water vapor so that there are only a few candidate sites on earth for CMB observation.
For now, all the existing ground-based CMB polarization experiments are located in southern hemisphere geographically such as the South Pole station in Antarctica, the Chajnantor observatory in Atacama Plateau of Chile. Some of these experiments have already worked out successfully in the past decade and some experiments are still in operation now.
In \cite{Ye}, the authors pointed out that Ali observatory, which is located in Tibetan Plateau of China, is a potential site for astronomical observation.

In this paper, we study in detail the observational conditions of Ali site.
Ali site is located at a peak in the west Gangdise mountain range next to the Himalayas. The Himalayas is known as the roof of the world, and will resist the wet air from the Indian Ocean to cross and reach Ali region. Along with the thin air due to the high altitude, Ali provides us ideal sites for CMB polarization observation. China's plan for CMB project includes two stages. The first stage is to build a multi-frequency telescope of 95/150 GHz at the 5250m site and the second one is to observe the CMB polarization at a higher site of about 6000m with more frequency bands.

This paper is organized as follows: Section II contains the detailed description of our study of Precipitable Water Vapor (PWV) and section III shows the observable sky and target field of AliCPT. The infrastructure of AliCPT-1 is briefly introduced in Section IV, then the last section is the conclusion.

\section{PWV calculation}

\begin{figure}
\begin{center}
\includegraphics[scale=0.45]{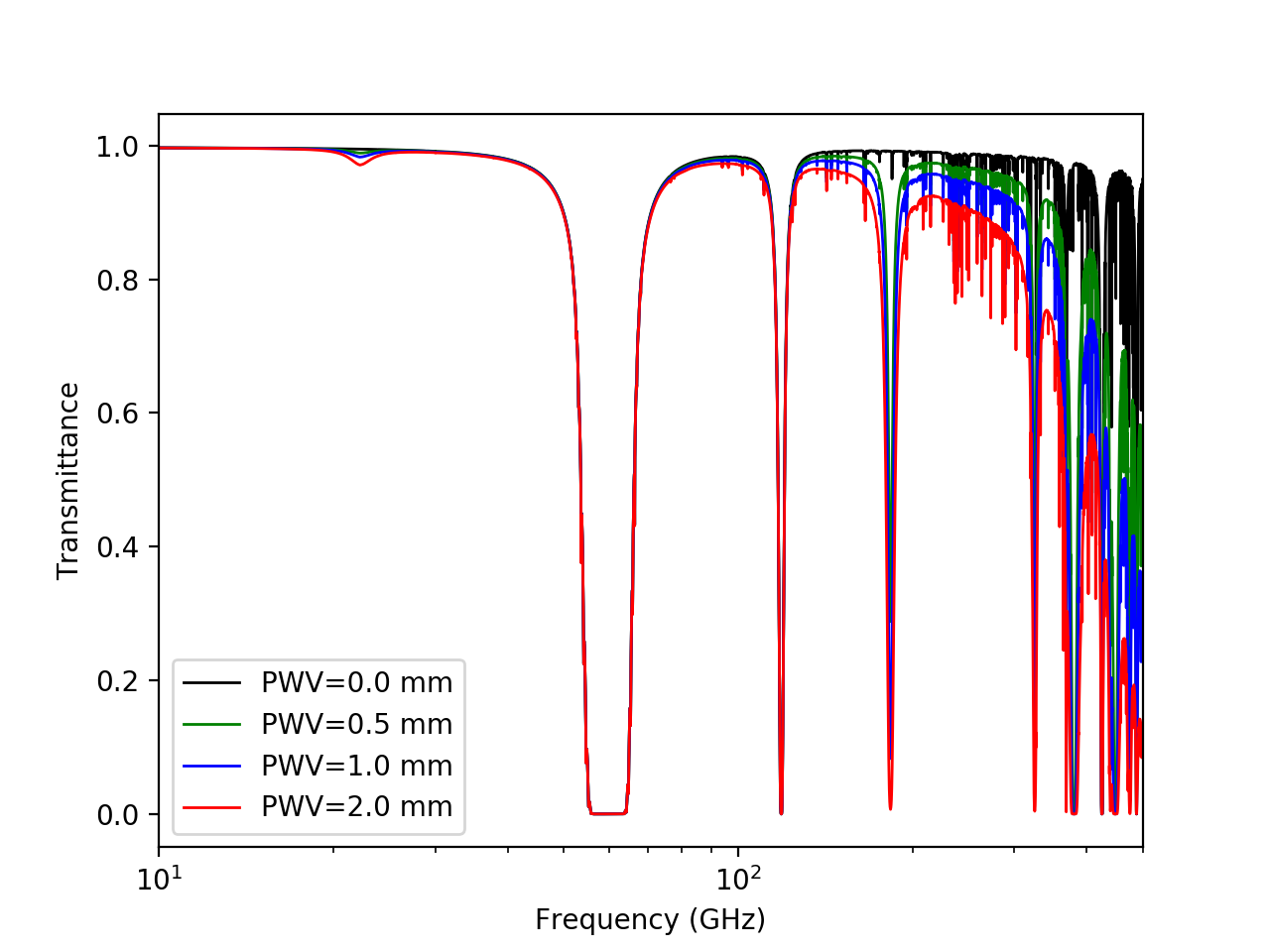}
\includegraphics[scale=0.45]{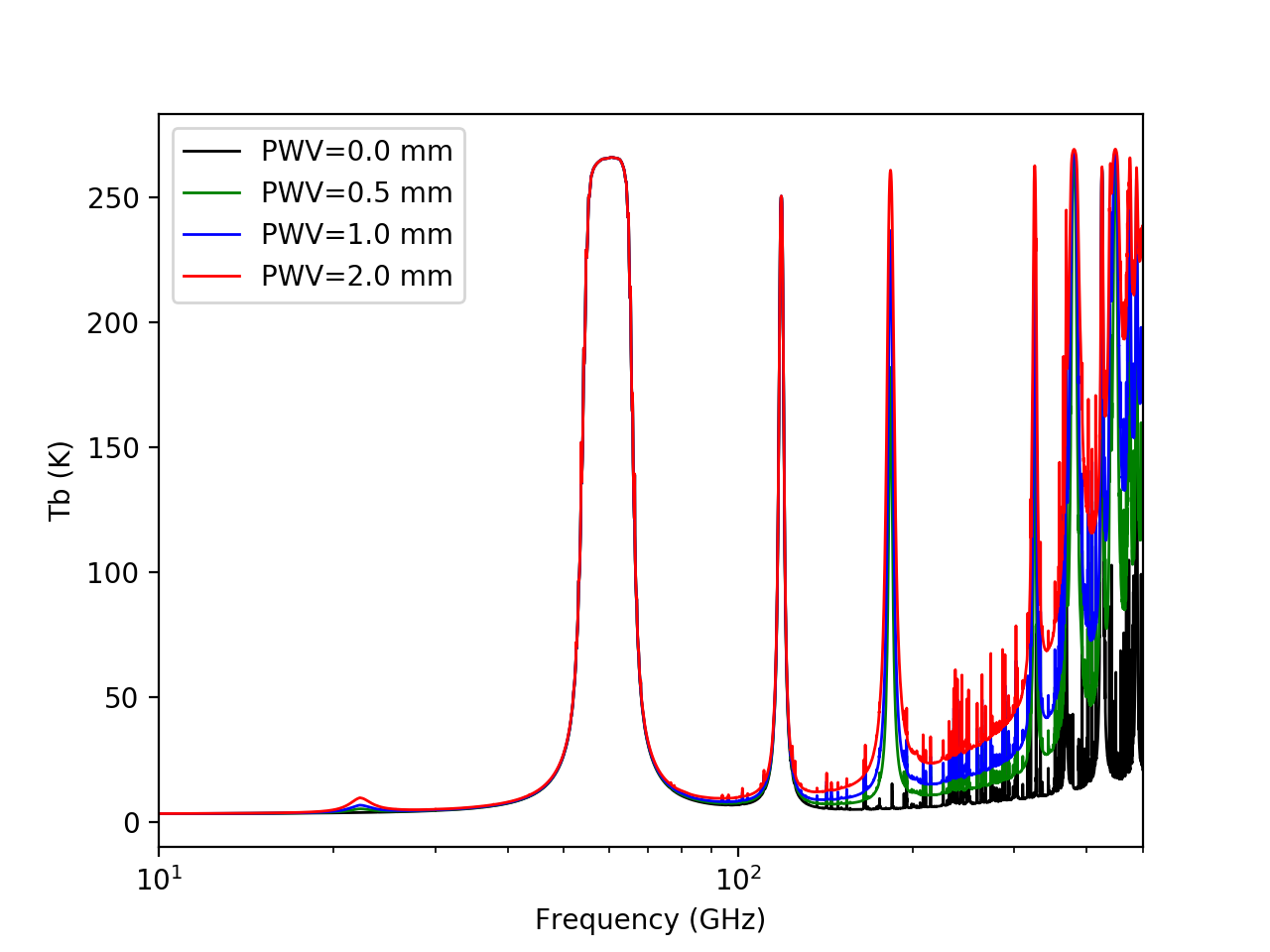}
\caption{Atmospheric transmittance (left) and brightness temperature (right) from the Ali observatory at the zenith for different PWV values ,obtained with the radiative transfer code $am$ \cite{Paine}}
\label{fig:atps}
\end{center}
\end{figure}

For ground-based CMB observations, the atmosphere will absorb and emit photons at millimeter and sub-millimeter wavelengths, reducing the significance of  the signals.  The strong absorption of different components of the air split the absorption spectrum into several windows at frequency 35, 90, 150, 220GHz and so on. The absorption can be characterized by the transmittance which is defined as the residual percentage of the incoming light through the atmosphere all the way down to the ground, and is shown in the left panel of Figure \ref{fig:atps}. In the right panel we use the brightness temperature at different frequencies as the intensity of the emission radiation. The absorption can introduce the attenuation of CMB signals, meanwhile, the emission can increase the number of photons and give rise to the photon noise. All these effects reduce the signal to noise ratio (SNR).

Among different components of the air, the water vapor is found to be the primary factor. PWV is the total amount of water vapor above unit cross-sectional area of the ground. We show three curves for different values of PWV (0.5, 1.0 and 2.0mm) in both panels. We find that at low frequency the transmission is high and the emission is low, and the changes caused by different PWV is tiny. Cases for high frequency are the opposite.
So for choosing an ideal location of CMB observation site, especially for higher frequencies, we need the atmosphere to be dry and stable around the site. The unique geographical conditions of Ali site can largely reduce the water vapor from the Indian Ocean. The average altitude for Ali region is more than 4200m, and there exists many mountain peaks higher than 5200m and even 6000m not very far from Ali observatory. All these factors make Ali a promising site for CMB observation.

In \cite{Ye}, two of us (Li and Zhang) with the other two authors (Ye, Su) derived 2-D distribution of annual median PWV with satellite reanalysis data, and found that Ali site has the potential for future CMB observations. However, their analysis is incomplete. In this paper, we focus on Ali as CMB observation sites. We study quantitatively water vapor condition for sites and show later that seasonal variation is crucial to the determination of the observing season. We will study in detail the atmospheric conditions for AliCPT-1 at 5250m and AliCPT-2 at 6000m and our result can be applied to at any altitude.

In our study, we use the MERRA-2 reanalysis data as well as the radiosondes data from the local Ali weather station. The Modern-Era Retrospective analysis for Research and Applications version 2 (MERRA-2) is the latest NASA/GMAO atmospheric reanalysis data produced using the Goddard Earth Observing System Model, Version 5 (GEOS-5) with its Atmospheric Data Assimilation System (ADAS), version 5.12.4. Here we use a 3 hours step time averaged data including temperature, height and specific humidity of 72 pressure levels, the time range is July 2011 to July 2017 and the resolution is $0.625^\circ (longitude) \times 0.5^\circ (latitude)$. MERRA-2 data has been validated by measurements at southern hemisphere such as the South Pole and Atacama\cite{Kuo}.
We choose (32.5N, 80.0E) as the reference point since it is closest to Ali sites $(32^\circ18'38''N,80^\circ01'50''E)$. We assume that the vertical atmospheric distribution of these two points are similar.

The radiosondes data is obtained from the local weather station of Ali Meteorological Service which is only a few kilometers away from the sites. Radiosondes is usually carried by a weather balloon and can measure various atmospheric parameters including altitude, temperature, pressure and humidity. Unless the cancellation caused by force majeure, the weather balloon is sent to the air twice a day (07:00 and 19:00). We apply almost the same method as for MERRA-2 data to calculate the PWV with the data from 2015 and 2016.

\begin{figure}
\begin{center}
\includegraphics[scale=0.35]{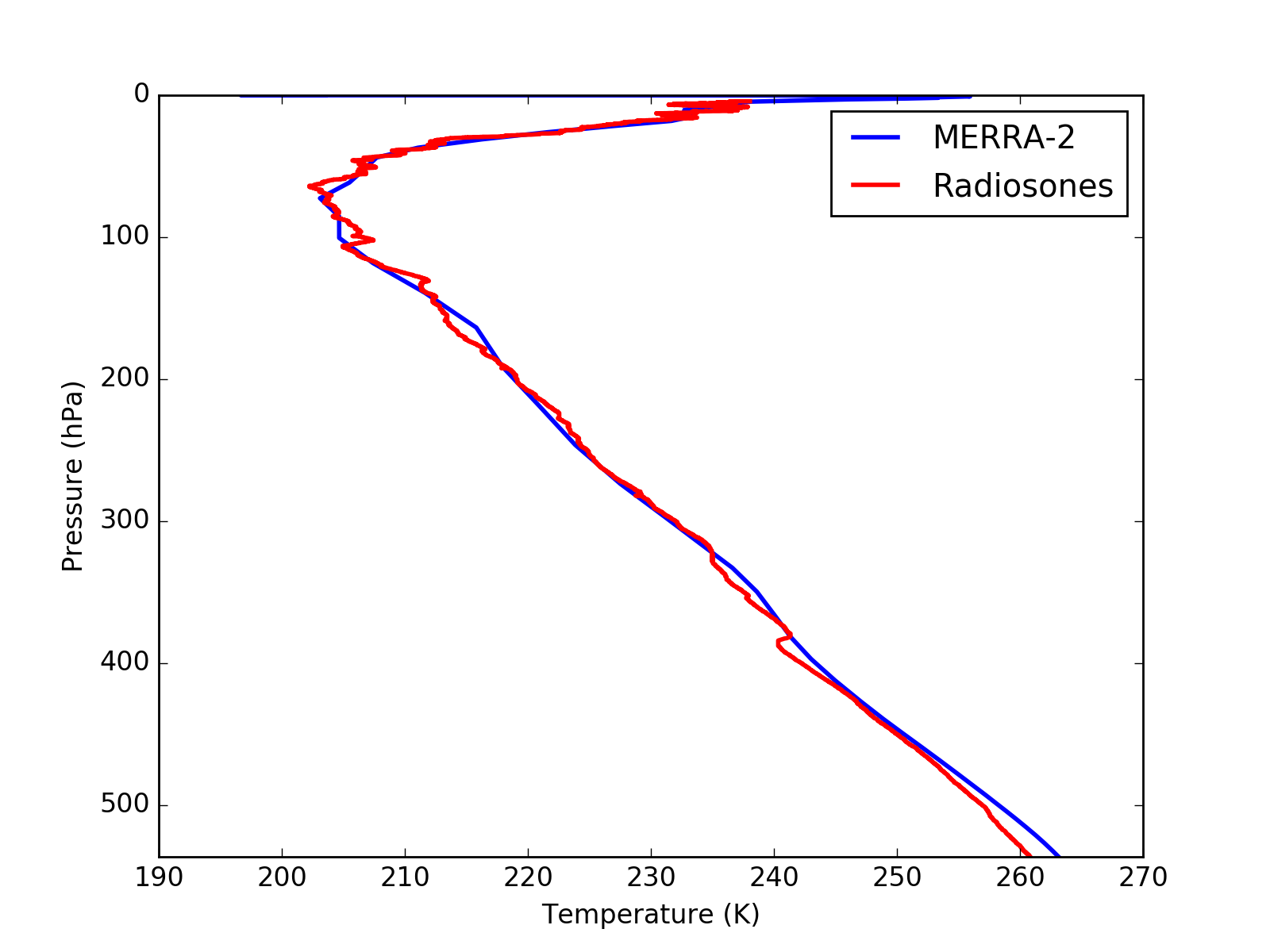}
\includegraphics[scale=0.35]{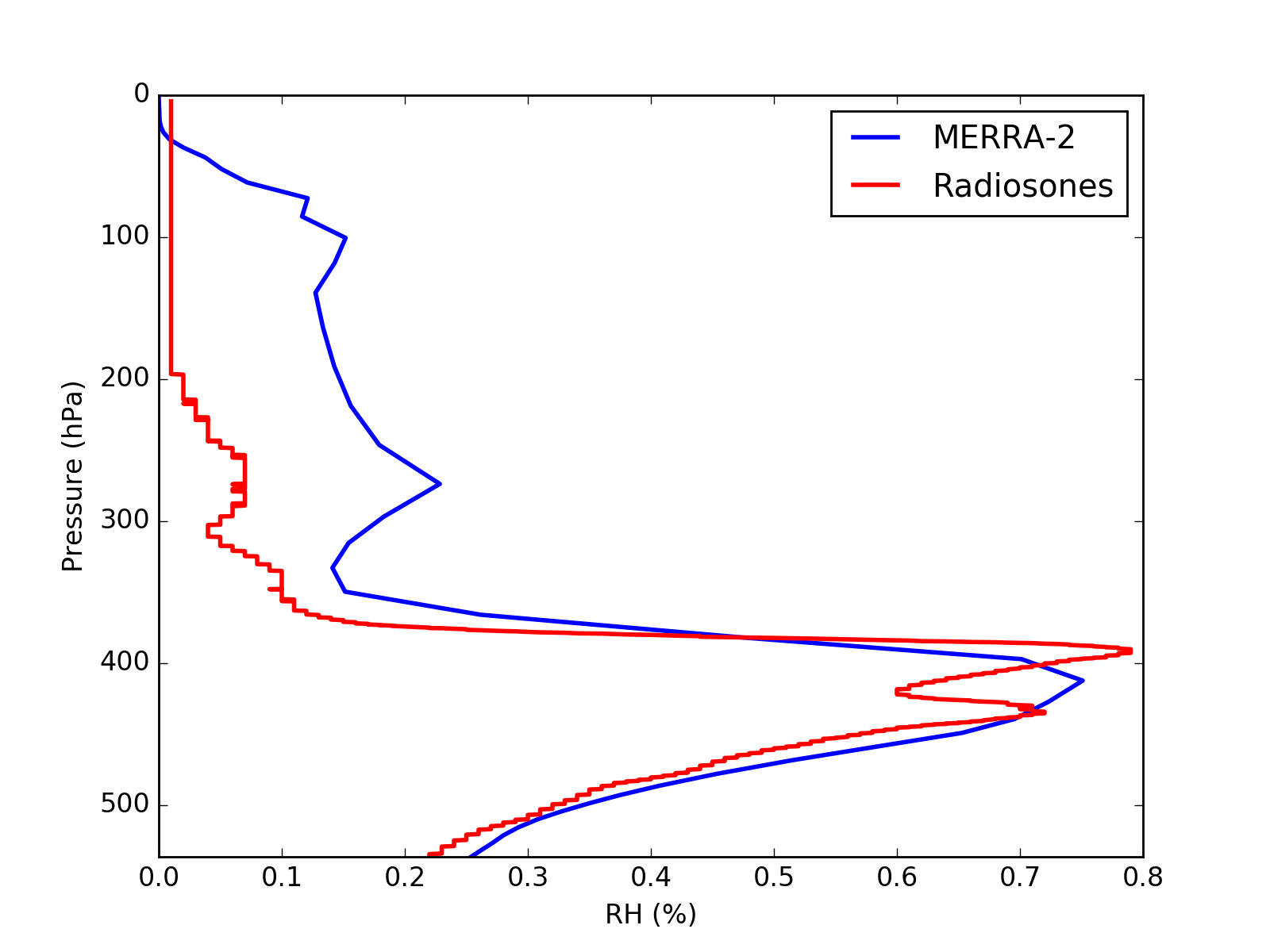}
\caption{Temperature data (left) and relative humidity data (right) from MERRA-2 reanalysis and radiosondes measurement for Ali site. The plots are obtained from data of one specific day.}
\label{fig:Temperature}
\end{center}
\end{figure}

Figure \ref{fig:Temperature} shows the comparison of the data obtained from MERRA-2 reanalysis and radiosondes measurement. The distribution of temperature over pressure shown in left panel are consistent with each other. For relative humidity in the right panel, we can see that these two data sets are consistent at low altitude. While at high altitude, relative humidity from radiosondes seems lower than result from MERRA-2. This may be the reason why the calculated PWV from radiosondes is usually smaller than that from MERRA-2 as described later in the results.

The formula we use to get the PWV is as follows:
\begin{align}
PWV=\int \rho_{\nu} dh = \int q_{\nu} \rho dh = -1/g \int q_{\nu} dp \approx -1/g \sum_{i}q_{\nu}^i \triangle p_i ~,
\end{align}
where $\rho_{\nu}$ is the density of water vapor, $q_{\nu}$ is the specific humidity, $p$ is for pressure, $g$ is the gravitational acceleration.

\begin{figure}
\begin{center}
\includegraphics[scale=0.35]{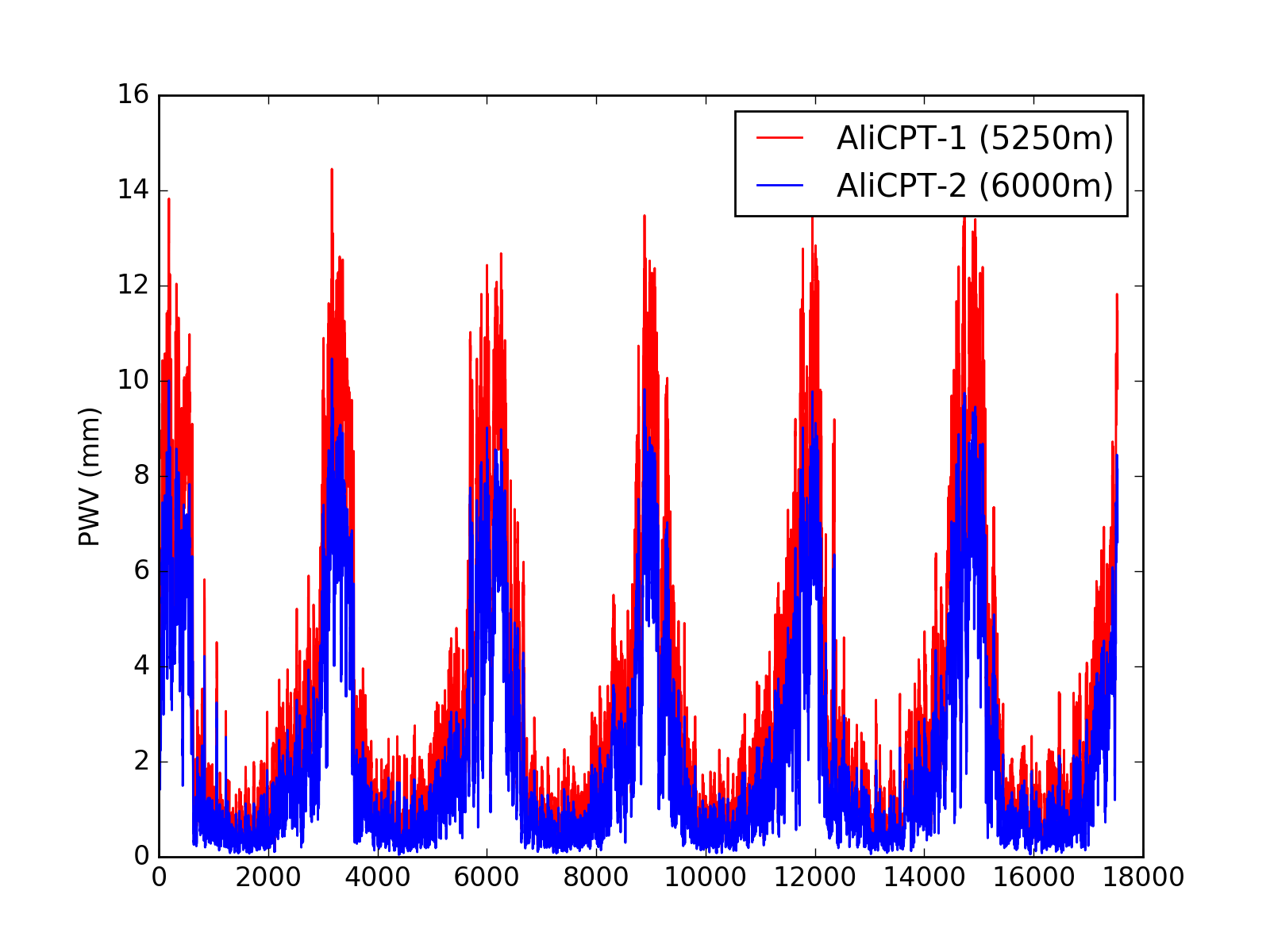}
\caption{PWV distribution for AliCPT-1 and AliCPT-2}
\label{fig:distpwv}
\end{center}
\end{figure}

Now we present our results. Firstly, we give the time distribution of the PWV over the whole 6 years in Figure \ref{fig:distpwv}. We can see that the amount of PWV in Ali has a heavy seasonal variation. The value during winter can be lower than 1mm while that in summer can be as large as 10mm. Besides, the PWV at 6000m is much smaller than that at the 5250m site. The large PWV in summer can be explained by the southerly monsoon, which brings huge amount of water vapor from the Bay of Bengal, the South China Sea and Western Pacific. However, in winter the water carried by mid-latitude Westerly wind from Arabian and Indian Ocean is strongly blocked by the Pamirs, the Himalayas and the Kunlun Montains \cite{Gai, Yao}.

\begin{figure}
\begin{center}
\includegraphics[scale=0.35]{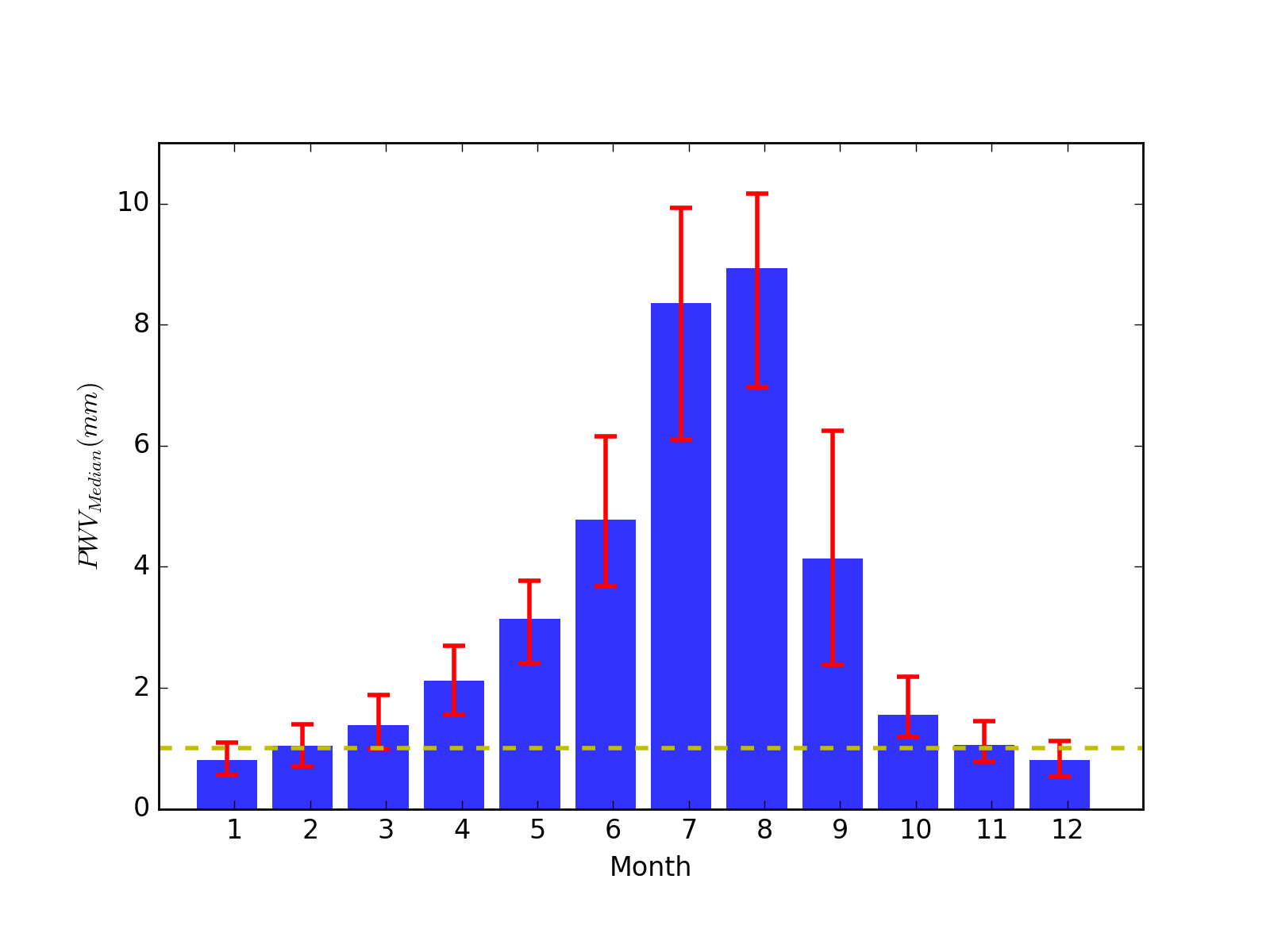}
\includegraphics[scale=0.35]{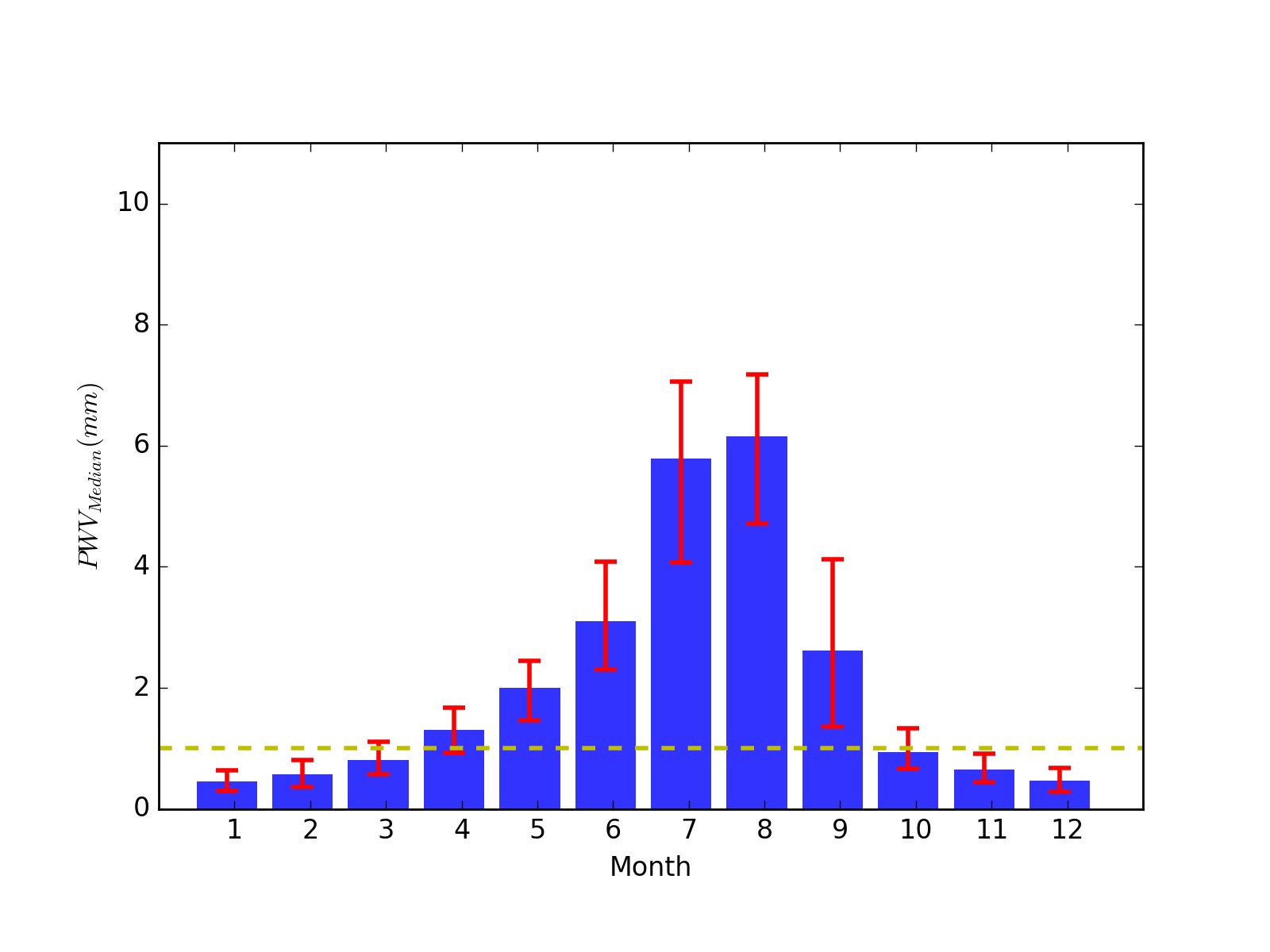}
\caption{Monthly distribution of PWV value averaged over 6 years (2011-2016) for AliCPT-1 site (left) and AliCPT-2 site (right)}
\label{fig:month_dist}
\end{center}
\end{figure}

To see the seasonal variation clearly, we plot the monthly quartile PWV amount of every month averaged over the 6 years in Figure \ref{fig:month_dist} (left) with the red bar as quartile points. The PWV in summer (June to August) is much higher than the rest months. In winter season, for 5250m site all the median values are about 1mm or below, which is good for CMB observation at 90/150GHz. About 25 percent time of March and October has a PWV below 1mm, so these two months should be included in the observing season to maximize the accumulate time of data. Thus, we choose October to March to be the observing season. For a future higher site at 6000m, the PWV is below 1.0mm for about 75\% of the five lowest months, the median value of PWV in winter can even be lower than 0.5mm, which may enable us to perform the observation at higher frequencies. The median PWV of choosing different months are summarized in Table \ref{tab:pwv}.

\begin{figure}
\begin{center}
\includegraphics[scale=0.35]{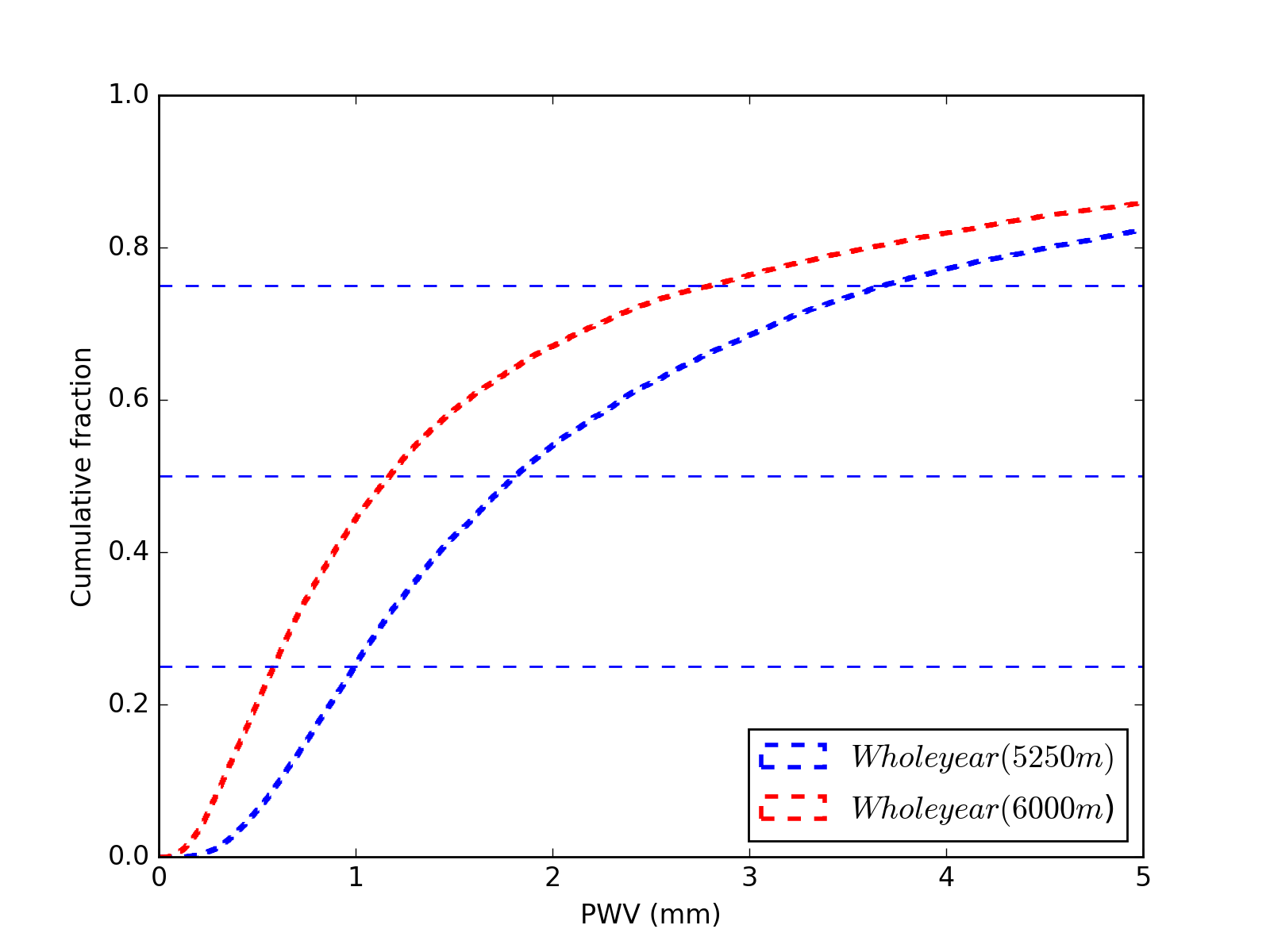}
\includegraphics[scale=0.35]{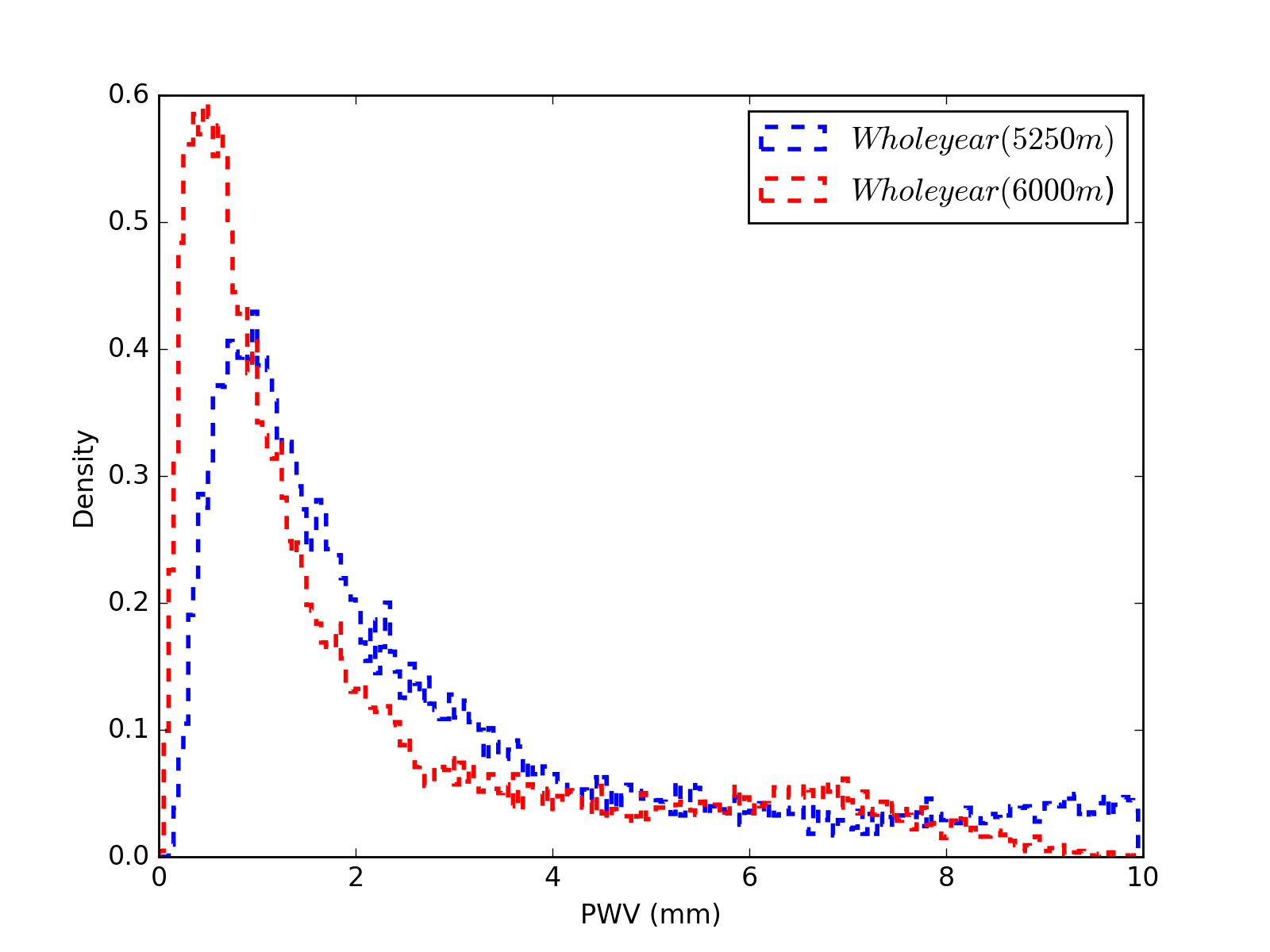}
\includegraphics[scale=0.35]{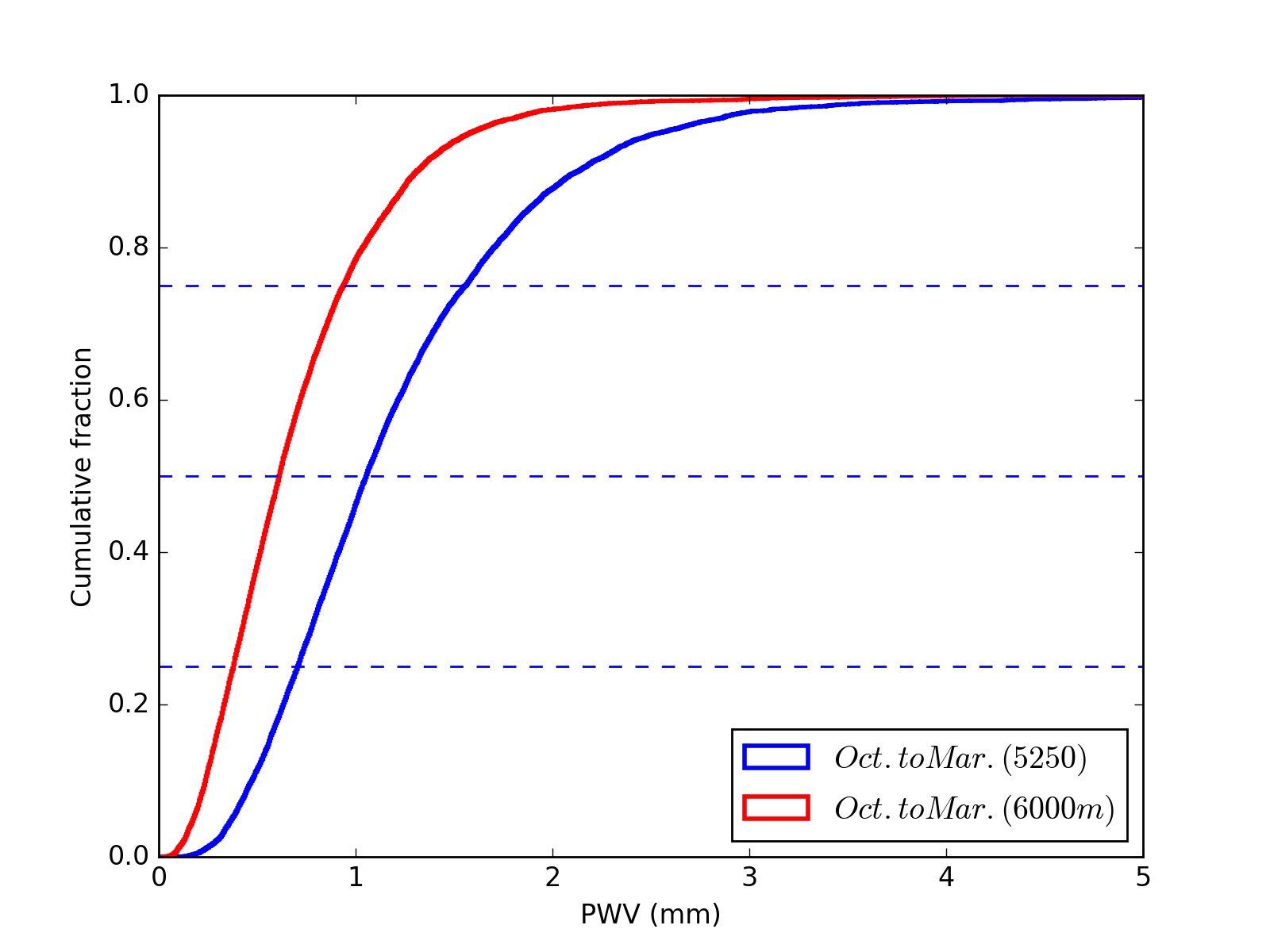}
\includegraphics[scale=0.35]{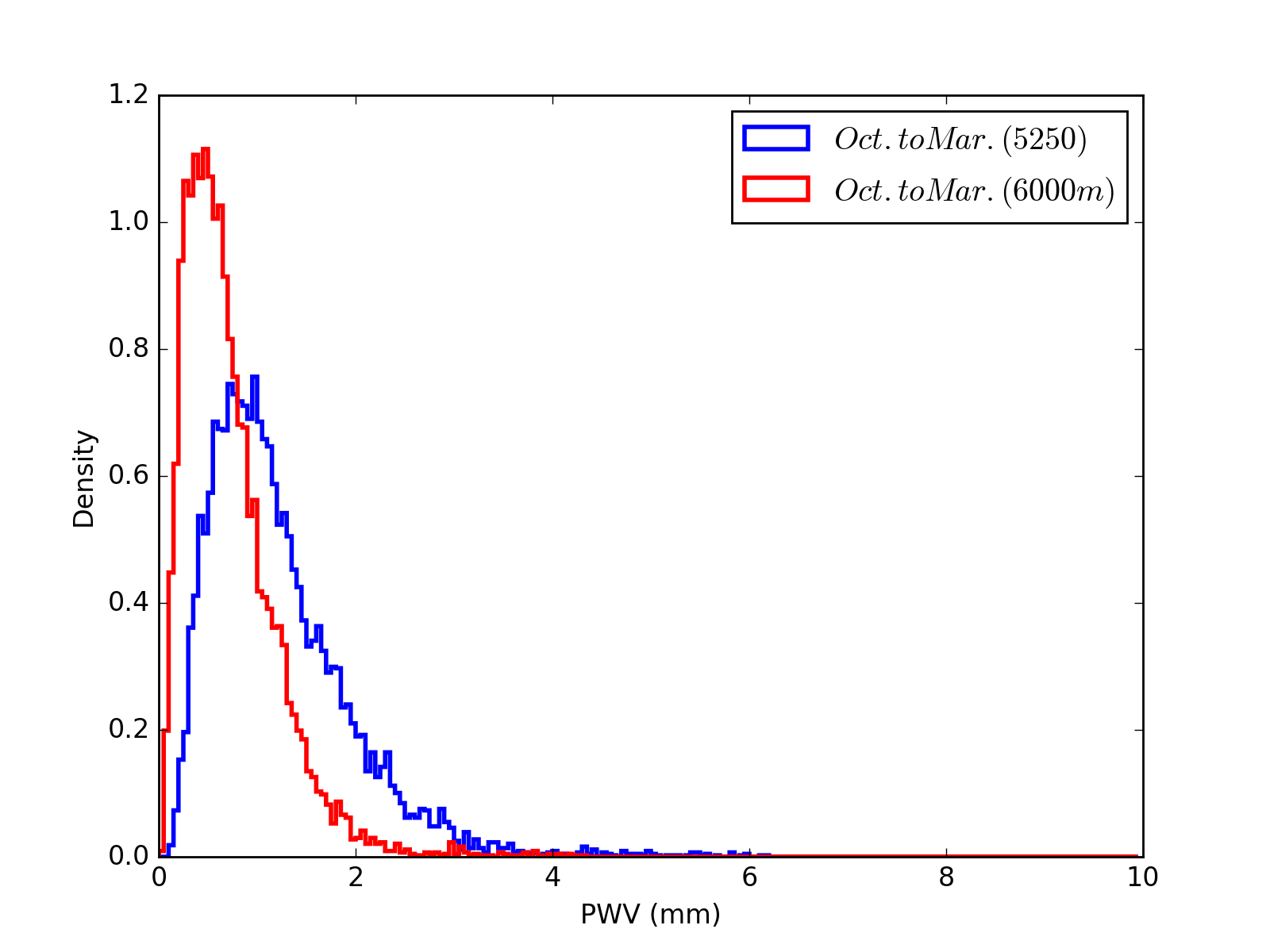}
\caption{The cumulative distribution (left column) and distribution density (right column) of calculated PWV from MERRA-2 reanalysis over 6 years for AliCPT-1 and AliCPT-2}
\label{fig:dist}
\end{center}
\end{figure}

We show the cumulative distribution of PWV (left column) and distribution density (right column) in Figure \ref{fig:dist}. The upper panels refer to the result of whole year and the lower panels are the results for the chosen observing season. The blue lines are for 5250m and the red lines are for 6000m. For the whole year statistics, the median PWV for the 5250m site is about 1.9mm and the PWV value with the highest probability density is about 1.0mm. These two quantities at 6000m are 1.2mm and 0.5mm. For the chosen season for observation, 90 percent  of PWV for the 5250 site are smaller than 2.0mm with a median value of about 1.1mm and 75 percent at 6000m are smaller than 1.0mm. The peak probability PWV is about the same as the whole year value. It's obvious that the higher site will be much better especially for higher frequency observation.

\begin{table}
\caption {Median PWV (mm) for the AliCPT with MERRA-2}  \label{tab:pwv}
\begin{center}
   \begin{tabular}{ c | c | c }
   \hline			
    Time range & AliCPT-1 (5250m) & AliCPT-2 (6000m)\\
   \hline
  Sep. - Apr. & 1.287 & 0.751\\
  Oct. - May. & 1.299 & 0.767\\
  Sep. - Mar. & 1.185 & 0.692\\
  Oct. - Apr. & 1.166 & 0.689\\
  Oct. - Mar. & 1.067 & 0.620\\
  Dec. - Feb. & 0.859 & 0.489\\
  \hline
  \end{tabular}
\end{center}
\end{table}

\begin{figure}
\begin{center}
\includegraphics[scale=0.35]{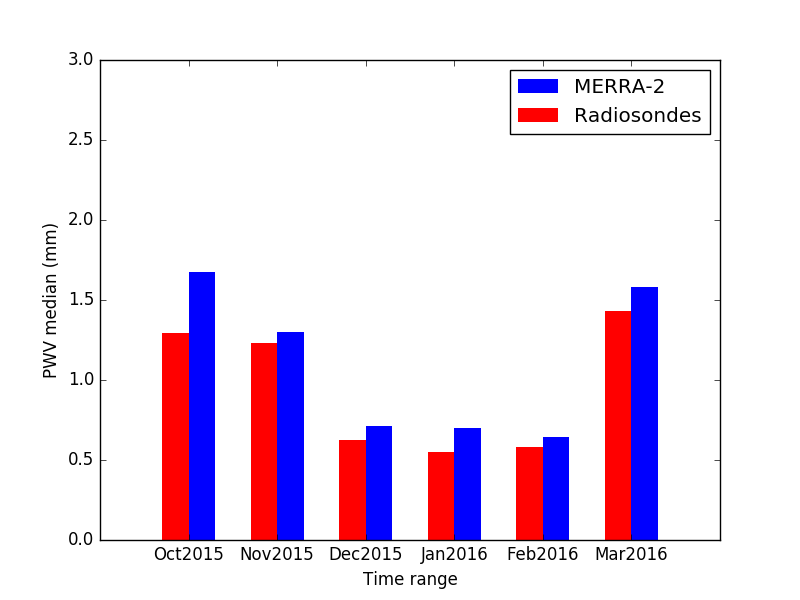}
\includegraphics[scale=0.35]{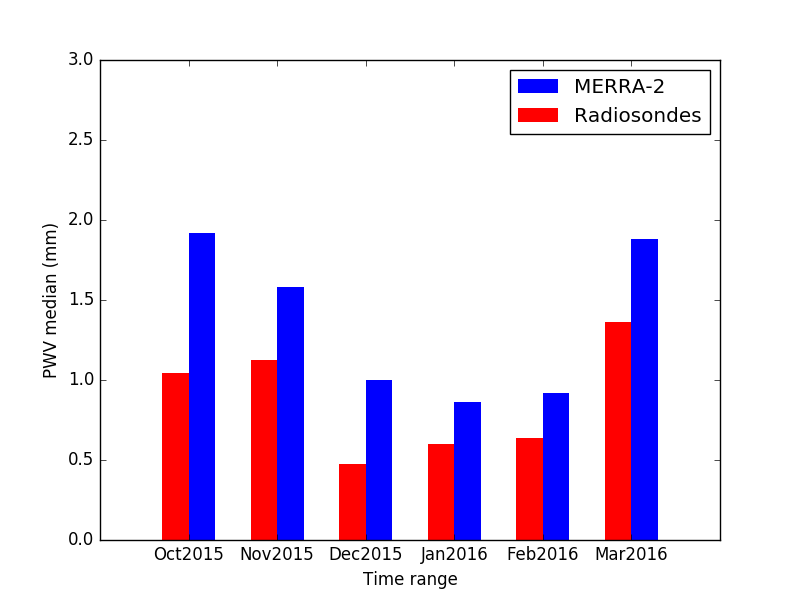}
\caption{The median PWV value for each month during a observing season (left for 07:00, right for 19:00)}
\label{fig:rs}
\end{center}
\end{figure}

\begin{figure}
\begin{center}
\includegraphics[scale=0.35]{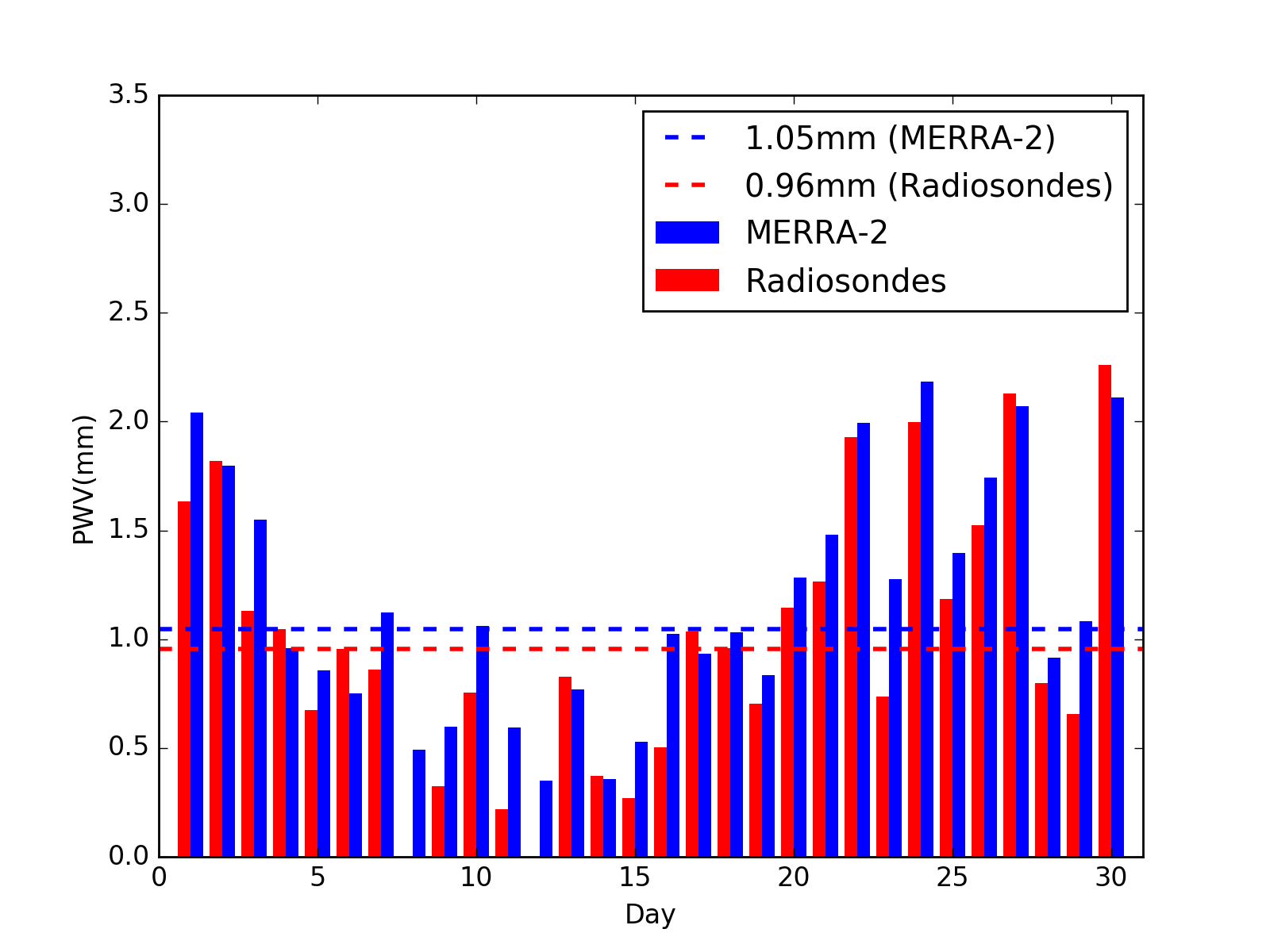}
\includegraphics[scale=0.35]{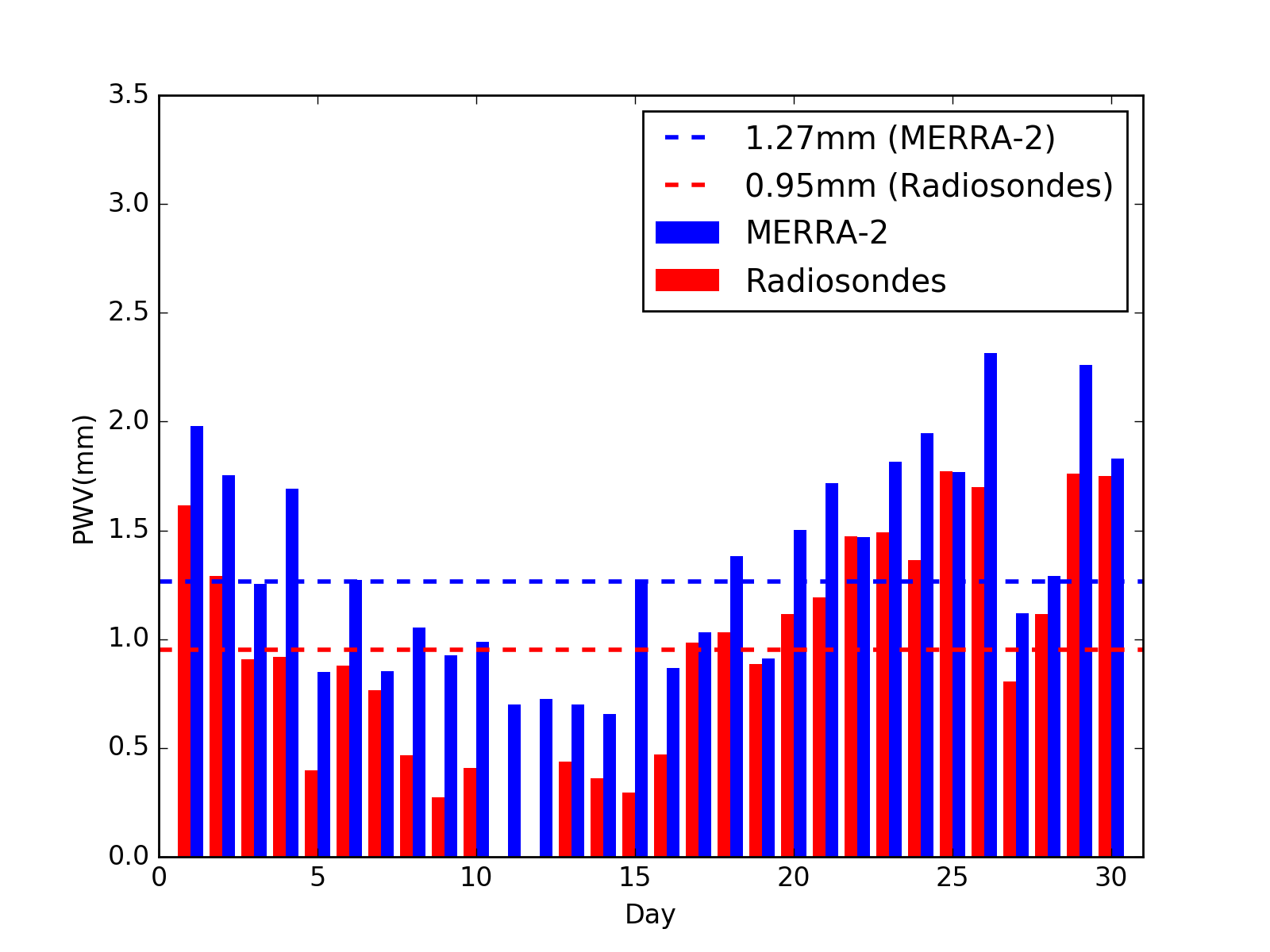}
\caption{The PWV for each day in Nov. 2016 calculated from radiosondes (red bar) and MERRA-2 (blue bar), left for 07:00 and right for 19:00. Blanks indicate no data provided by radiosondes. }
\label{fig:r_vs_s}
\end{center}
\end{figure}

As a comparison, we plot the histogram of the median PWV of the observing season with both MERRA-2 and radiosonde data in Figure \ref{fig:rs}. We can see that both panels show the same variation tendency. In this sense, result from these two datasets are consistent with each other. We can also see that the MERRA-2 result is slightly larger than that from the radiosondes, and this is more obvious in the 19:00 result. To understand this feature, we show the PWV of a specific month from both datasets in Figure \ref{fig:r_vs_s}. This figure has the same feature as Figure \ref{fig:rs}. These numbers are summarized in Table \ref{tab:pwv}.

\begin{figure}
\begin{center}
\includegraphics[scale=0.35]{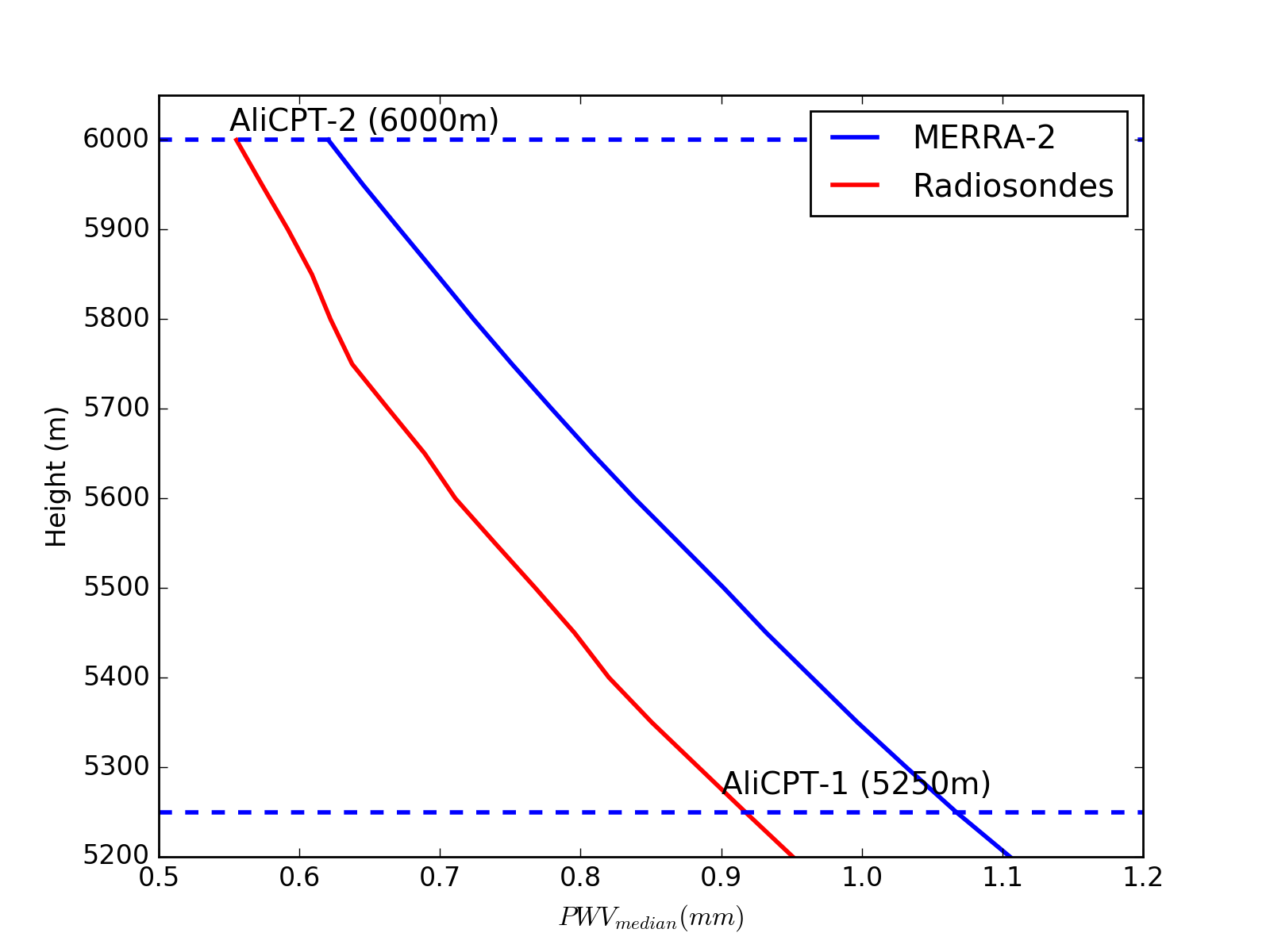}
\includegraphics[scale=0.35]{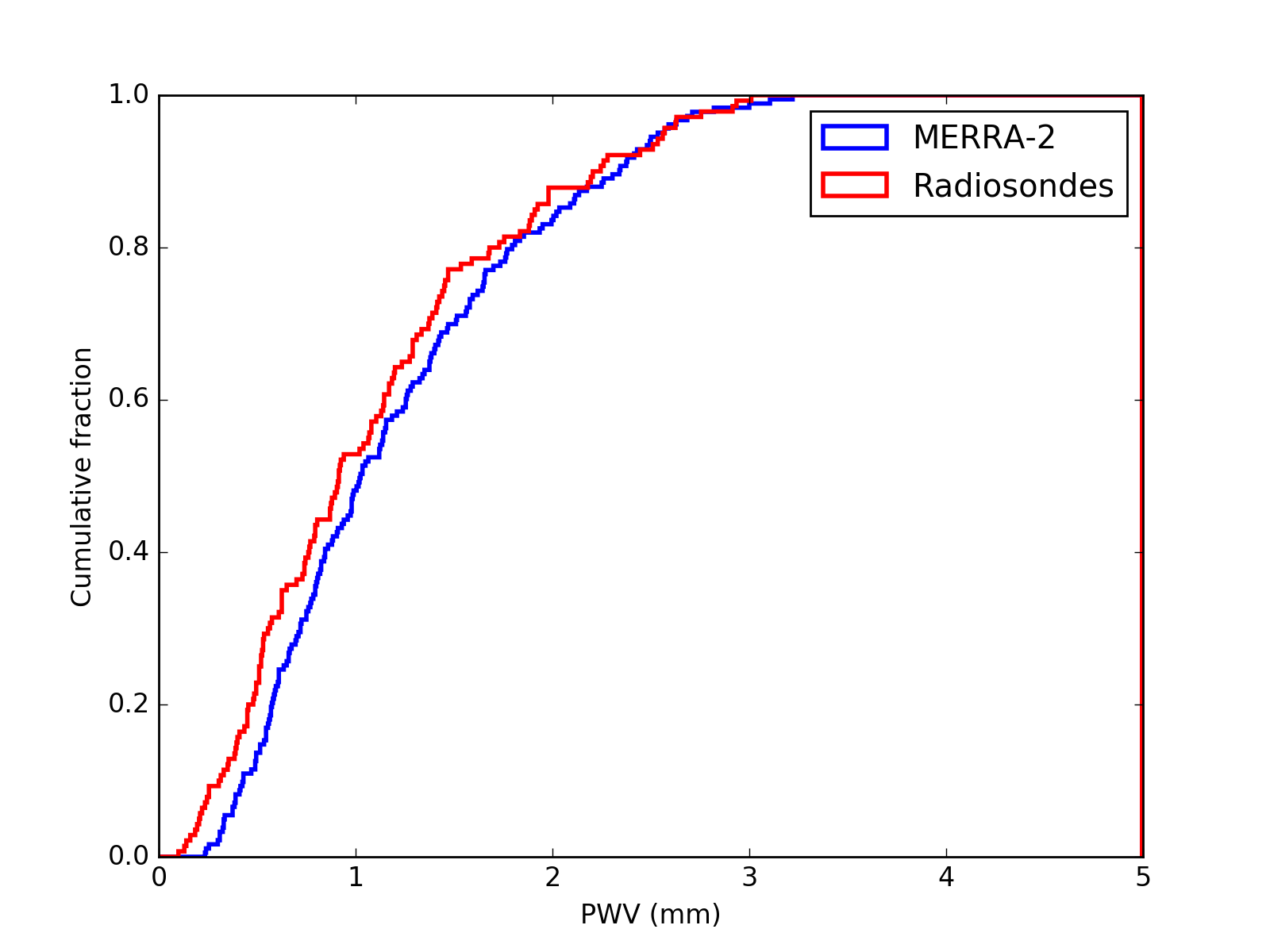}
\caption{Comparison between the MARRA-2 and radiosonde result: PWV VS height for Ali site in the left and Cumulative fraction in the right}
\label{fig:comparison}
\end{center}
\end{figure}

We also plot the PWV versus height curves of the two datasets in the left panel of Figure \ref{fig:comparison} and cumulative fraction of the chosen season in the right panel to show the difference between results from the two datasets explicitly. The MERRA-2 data is discrete in 2-D distribution on the ground, this may lead to the difference between a local measurement and a grid sample, since the terrain around the area is very steep \cite{Ye, Suen}. The comparison between MERRA-2 and site survey in Atacama Desert and South Pole has been done in Kuo's paper \cite{Kuo}. In his paper, he got the same conclusion, that PWV from MERRA-2 are comparable with the on-site measurement and the former is slightly higher.

As the conclusion of this section, we used two different meteorological atmospheric datasets (satellite and radiosondes) to analyze the PWV value of the 5250m AliCPT-1 site as well as 6000m height for AliCPT-2. We found that MERRA-2 is in good agreement with the radiosondes. It's appropriate to choose October to March as the observing season. For the 5250m site, we found the median PWV during the chosen season is about 1mm (1.07mm for MERRA-2, 0.92mm for radiosondes), which is excellent to deploy AliCPT-1 at the current chosen site. For 6000m height, The median PWV is about 0.6mm (0.62mm for MERRA-2, 0.56mm for radiosondes), this is even better for CMB observation especially for higher frequencies.

\section{Observable sky and target field}
After \emph{COBE (Cosmic Background Explorer)} first measured the anisotropies of the CMB temperature field, satellite projects like \emph{WMAP (Wilkinson Microwave Anisotropy Probe)} and \emph{Planck} gave us more precise results of the temperature field of CMB and its power spectrum. However, the measurement of polarization field and the corresponding power spectra are still not precise enough, especially for B-mode polarization. Ground-based experiments can provide a better measurement of polarization, but only partial sky is observable for one experiment. BICEP at the south pole can cover a part of the southern sky and POLARBEAR at Atacama desert of Chile have the ability to cover almost the whole southern sky and also have access of a part of northern sky. To be able to cover the whole sky on the ground, one more project in northern hemisphere is needed.

\begin{figure}
\includegraphics[scale=0.5]{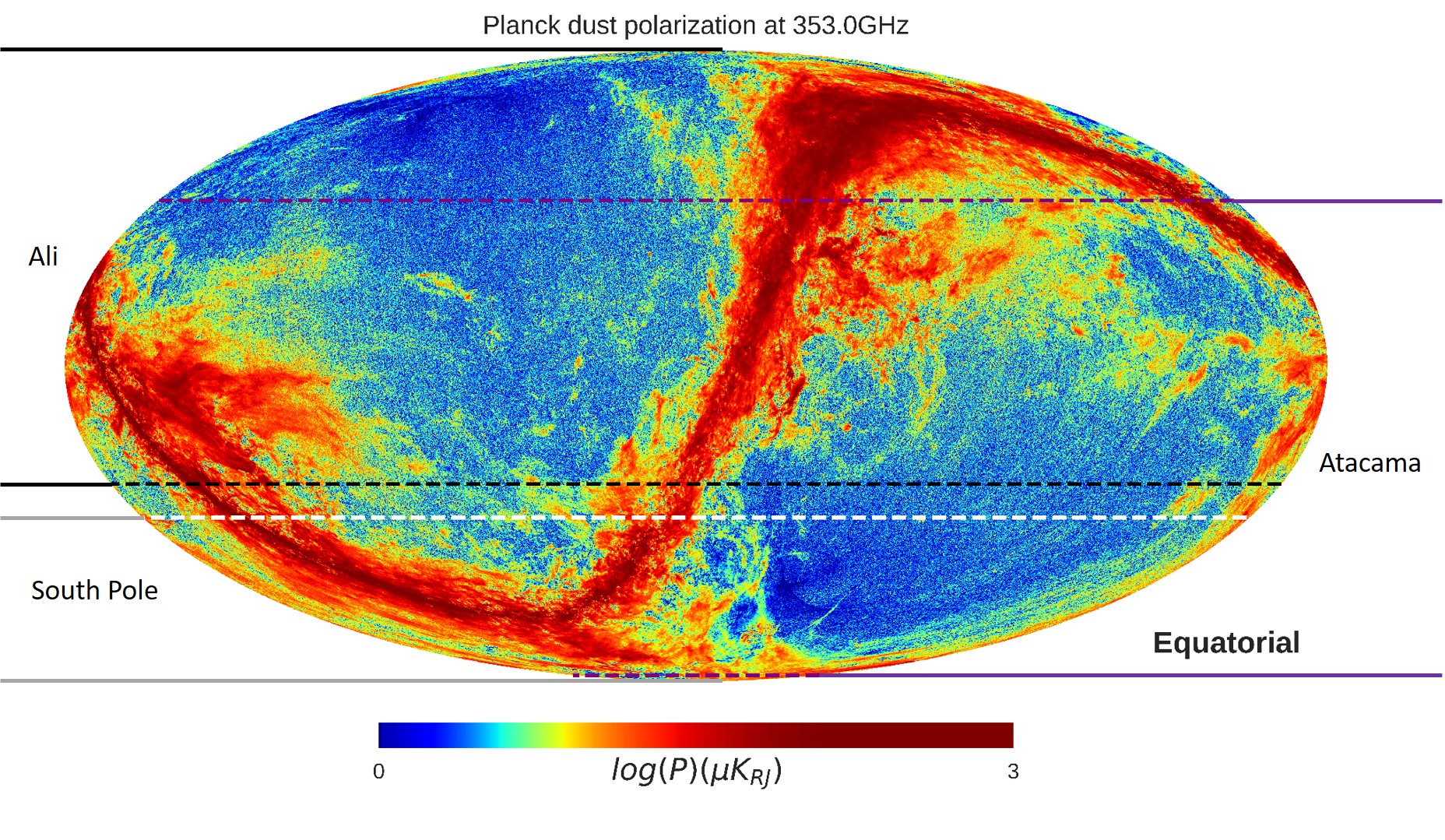}
\caption{Plots of the observable sky for AliCPT together with BICEP (the South Pole) and POLARBEAR (Atacama). Region above the black dashed line is for AliCPT, regions below the white and purple dashed lines are for BICEP and POLARBEAR. The background is the dust polarization brightness temperature from Planck 2015 results. The minimum and maximum are set to be $1.0 \mu K_{RJ}$ and $1.0\times10^3 \mu K_{RJ}$}\label{observable_sky}
\end{figure}

As far as we know, the PGWs can only be effectively detected by measuring the B-mode polarization of CMB. However, the galactic foreground consisting of many astrophysical diffuse components such as dust and synchrotron, they turned out to be the major problems to extract the signal. So it's important to target at the sky regions with less foreground.

AliCPT are located in the mid-latitude region of the northern hemisphere. It can cover the whole northern sky and also a part of the southern sky.
To have a quantitatively analysis, we take a set of instrumental parameters, a three-axis driving (azimuth, elevation and boresight) mount with a $45^\circ$ lowest elevation and a $\sim30^\circ$ field of view. As shown in figure \ref{observable_sky}, Ali observable sky is above the black dashed lines. The observable sky of the south pole (BICEP) and Atacama (POLARBEAR) are areas below the the white and purple dashed lines.
For BICEP and POLARBEAR, we used site and instrumental parameters from reference \cite{Yoon:2006jc}\cite{Ahmed:2014ixy}\cite{Stebor:2016hgt}. The background is the brightness temperature of the dust polarization at reference frequency 353.0 GHz from Planck 2015 result, and we set the minimum and maximum to be $1.0 \mu K_{RJ}$ and $1.0\times10^3 \mu K_{RJ}$. We can see that a whole sky coverage can be realized by the combination of Ali, the South Pole and Atacama.

\begin{figure}
\includegraphics[scale=0.5]{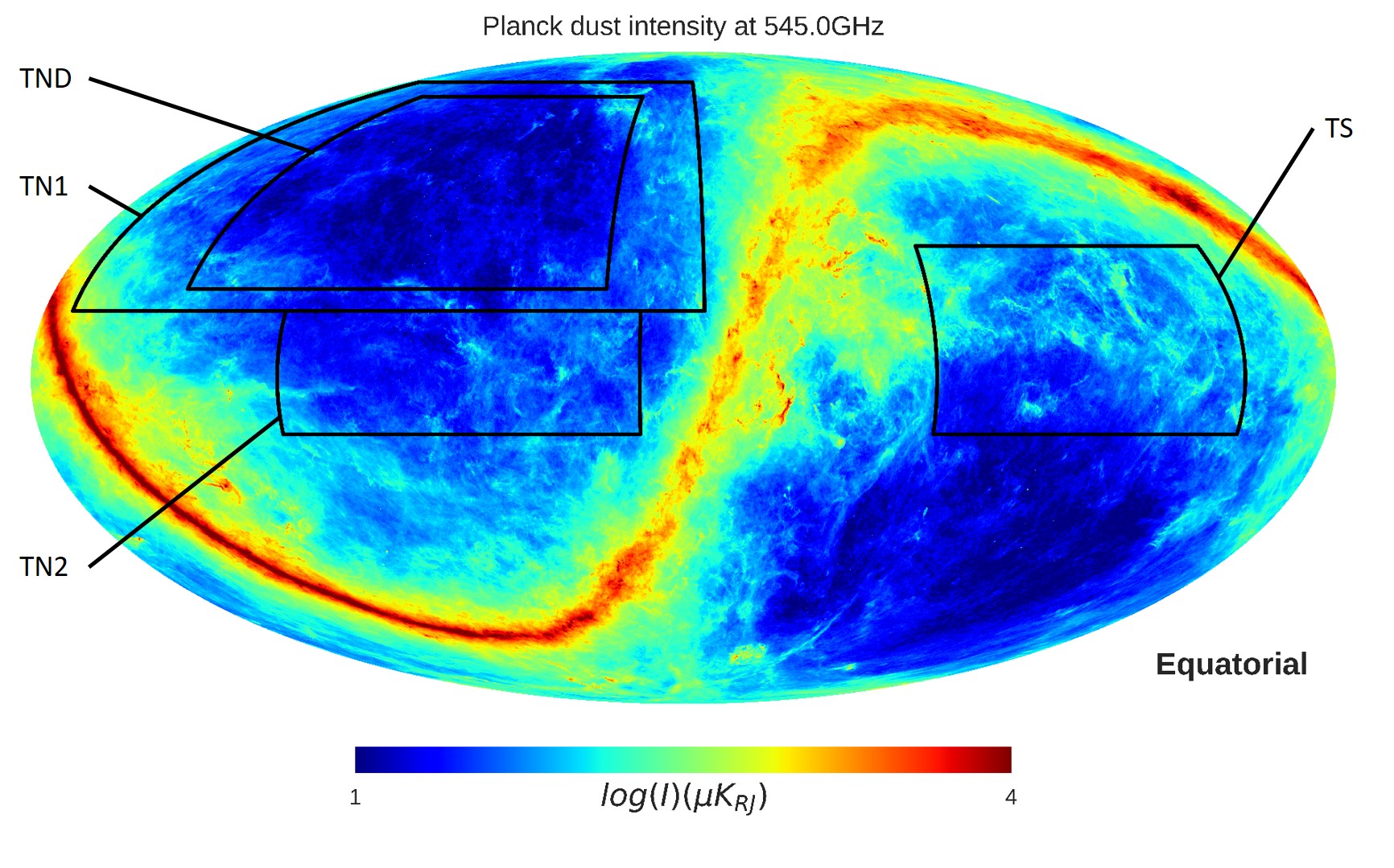}
\caption{Target fields for AliCPT. TN1 and TN2 regions are the target fields in the northern galactic hemisphere, TS is the target field in the southern galactic hemisphere. The smaller TND region with the lowest foreground is for deeper scan. The background is the dust intensity brightness temperature from Planck 2015 results. The minimum and maximum are set to be $10.0 \mu K_{RJ}$ and $1\times10^4 \mu K_{RJ}$}\label{target_field}
\end{figure}

For a B-mode polarization measurement, we need to survey the low dust emission region, which is shown in figure \ref{target_field} as the dark blue area. The red area is the dirty area and is coincide with the milky way.  Ali can cover almost all the clean areas in the northern galactic hemisphere and a part of the clean areas in the southern galactic hemisphere, which is shown in figure \ref{target_field} as region TN1,TN2 and TS.
We use the brightness temperature of the intensity of dust at 545.0 GHz from Planck 2015 result\cite{Adam:2015wua} as the background and set the range to be $10.0 \mu K_{RJ} \sim 1.0\times10^4 \mu K_{RJ}$.
Atacama is also able to cover the lower latitude northern sky region. However, the weather condition is not ideal for most time of the observing season when the clean region is accessible from Atacama\cite{Kuo}. The overlapped region of AliCPT and POLARBEAR also make it convenient to do crosscheck and cross-correlation studies.

To summarize, Ali is a perfect site for CMB observation in the north. The observable sky and observable low dust intensity region are complement to that of the running projects at the South Pole and Atacama. That will be very important for cosmological studies, especially for the detection of the PGWs.

\section{Infrastructure}
The infrastructure is also a key factor for CMB experiment. The AliCPT-1 site is only 1km away from the Ali observatory. In Ali observatory some optical telescopes have been built already, the Internet is provided and electric power is connected to the city grid. There is a daily commercial flight from Ali to Lhasa, the capital of Tibet. It takes about 30 minutes to drive from the site to airport. Also, the capital of Ali region, Shiquanhe town, with a population of about 20,000 is less than 30 minutes drive to the AliCPT-1 site.
The construction of AliCPT-1 site started in March of 2017 and is to be finished at the end of this year. The observation is expected to start in 2020.

\section{Conclusions}
In this paper, we have studied the profile of the AliCPT sites in atmospheric conditions and sky coverage.
Our results show that the 6 months median PWV for AliCPT-1 site is about 1.0mm, which is good for the planned CMB observation at 90/150GHz. For AliCPT-2 with a higher altitude, the PWV is about 0.6mm and is promising for higher frequencies observation. Moreover, due to its mid-latitude location, Ali can cover the whole northern sky and partial southern sky. The northern galactic sky with the lowest foreground contamination is within the observable area of Ali sites. We have also calculated the Liquid Water Path (LWP) and Ice Water Path (IWP) for Ali, and found that the results are consistent with \cite{Kuo}.

The main focus of this paper is on the site condition of AliCPT, however, we also evaluate other sites as shown in Table \ref{tab:site}. As only the atmospheric environment is concerned, the best candidate site is Dome-A. But the infrastructure of Dome-A is not well-developed for now. When it comes to sky coverage, all three sites in the southern hemisphere is difficult to access northern sky.
Considering all aspects including the atmospheric condition, sky coverage together with the infrastructure, we come to the conclusion that Ali offers excellent CMB observation sites in northern hemisphere.

\begin{table}
\caption {Profile for different sites. Labels $^1$ and $^2$ represent the PWV obtained with MERRA-2 and radiosondes.}
\label{tab:site}
\begin{center}
   \begin{tabular}{ c  | c | c | c | c | c  }
   \hline			
    Site  & Height(m) & Time range & PWV(mm) & Sky range & Observable fraction (\%) \\
   \hline
 $ \text {AliCPT(5250)}$  & 5250 & Oct. - Mar. & 1.07 & Whole North + Part South & 70\\
  \hline
 $ \text {AliCPT(6000)}$ & 6000 & Oct. - Mar. & 0.62 & Whole North + Part South & 70\\
  \hline
  South Pole(BICEP3) &2835 & Apr. - Sep. & 0.27 & Part South & 20\\
  \hline
  Atacama(POLAEBEAR) & 5200 & Apr. - Sep. & 0.85 & Whole South + Part North & 80\\
  \hline
  Dome A & 4093 & Apr. - Sep. & 0.12 & Part South & 25\\
  \hline

  \end{tabular}
\end{center}
\end{table}

\section{Acknowledgements}
We thank Chao-lin Kuo, Cong-Zhan Liu, Di Wu, Fang-Jun Lu, He Xu, Hong-Shuai Wang, Li-Yong Liu, Yong-Qiang Yao, Yong-Jie Zhang, Yun-He Zhou and Zheng-Wei Li for useful discussions. Si-Yu Li thanks Zhen Bai of Ali Meteorological Service for providing the radiosondes data.

The work of Yong-Ping Li, Yang Liu, Si-Yu Li and Xinmin Zhang are supported by the National Science Foundation of China (NSFC) under Grant No. 11375202. Hong Li is supported by the Youth Innovation Promotion Association Project of CAS. This work is also supported in part by NSFC under Grant No. 11653001, Pilot B Project of CAS (No. XDB23020000) and Sino US Cooperation Project of Ministry of Science and Technology (No. 2016YFE0104700).


\begin{thebibliography}{999}

\bibitem{Seljak:1996gy}
  U.~Seljak and M.~Zaldarriaga,
  Phys.\ Rev.\ Lett.\  {\bf 78}, 2054 (1997),
  astro-ph/9609169.

\bibitem{Kamionkowski:1996zd}
  M.~Kamionkowski, A.~Kosowsky and A.~Stebbins,
  Phys.\ Rev.\ Lett.\  {\bf 78}, 2058 (1997),
  astro-ph/9609132.

\bibitem{Li:2008tma}
  M.~Li and X.~Zhang,
  Phys.\ Rev.\ D {\bf 78}, 103516 (2008),
  arXiv:0810.0403 [astro-ph].

\bibitem{Zhao:2014yna}
  W.~Zhao and M.~Li,
  Phys.\ Rev.\ D {\bf 89}, no. 10, 103518 (2014),
  arXiv:1403.3997 [astro-ph.CO].

\bibitem{Zhao:2015mqa}
  G.~B.~Zhao, Y.~Wang, J.~Q.~Xia, M.~Li and X.~Zhang,
  JCAP {\bf 1507}, no. 07, 032 (2015),
  arXiv:1504.04507 [astro-ph.CO].

\bibitem{Li:2015vea}
  S.~Y.~Li, J.~Q.~Xia, M.~Li, H.~Li and X.~Zhang,
  Phys.\ Lett.\ B {\bf 751}, 579 (2015),
  arXiv:1506.03526 [astro-ph.CO].

\bibitem{Ye}
  Q.~Z.~Ye, M.~Su, H.~Li, et al.,
  Mon. Not. Roy. Astron. Soc. Lett. 2015, {\bf 457}, 1,
  arXiv:1512.01099 [astro-ph.IM].



\bibitem{Suen}
  J.~Suen, M.~Fang and P.~Lubin,
  IEEE Transactions on Terahertz Science and Technology. vol.4, pp.86 - 100, 2014.

\bibitem{Paine}
  S.~Paine, 
  Submillimeter Array Technical Memo \#152.

\bibitem{Kuo}
  C.~L.~Kuo,
  Astrophys.\ J.\  {\bf 848} (2017) no.1,  64,
  arXiv:1707.08400 [astro-ph.IM].

\bibitem{MERRA}
M.~G.~Bosilovich {\it et al.}, 
NASA Technical Report Series on Global Modeling and Data Assimilation,
Volume 43, 2015.

\bibitem{Gai}
  G.~Xue, Z.~Lin and Q.~De,
  Sci. Tech. Engrg, 2013, 13: No.35.

\bibitem{Yao}
  Y.~Yao, X.~Lei, L.~Zhang, B.~Zhang, H.~Peng and J.~Zhang,
  Chin Sci Bull, 2016, 61: 1462-1477.

\bibitem{Yoon:2006jc}
  K.~W.~Yoon {\it et al.},
  Proc.\ SPIE Int.\ Soc.\ Opt.\ Eng.\  {\bf 6275}, 1K (2006),
  astro-ph/0606278.

\bibitem{Ahmed:2014ixy}
  Z.~Ahmed {\it et al.} [BICEP3 Collaboration],
  Proc.\ SPIE Int.\ Soc.\ Opt.\ Eng.\  {\bf 9153}, 91531N (2014),
  arXiv:1407.5928 [astro-ph.IM].

\bibitem{Stebor:2016hgt}
  N.~Stebor {\it et al.},
  Proc.\ SPIE Int.\ Soc.\ Opt.\ Eng.\  {\bf 9914} (2016) 99141H.

\bibitem{Adam:2015wua}
  R.~Adam {\it et al.} [Planck Collaboration],
  Astron.\ Astrophys.\  {\bf 594}, A10 (2016),
  arXiv:1502.01588 [astro-ph.CO].

\bibitem{Adam:2014bub}
  R.~Adam {\it et al.} [Planck Collaboration],
  Astron.\ Astrophys.\  {\bf 586}, A133 (2016),
  arXiv:1409.5738 [astro-ph.CO].

\end{thebibliography}
\end{document}